\documentclass{pasa}%

\title[Torus Instability and QPOs]{Angular Velocity Perturbations Inducing
  the Papaloizou-Pringle Instability and QPOs in the Torus around the Black Hole}

\author[O. Donmez]{Orhan Donmez
  \thanks{orhan.donmez@aum.edu.kw}\\
\affil{College of Engineering and Technology, American
  University of the Middle East (AUM), Egaila, Kuwait}}%
\jid{PASA}
\doi{10.1017/pas.\the\year.xxx}
\jyear{\the\year}

\usepackage[authoryear]{natbib}
\bibpunct{(}{)}{;}{a}{}{,}
\usepackage{graphicx}
\usepackage{amsmath}

\usepackage{epsfig}

\usepackage{aas_macros}
\usepackage{hyperref} 
\hypersetup{colorlinks,citecolor=blue,linkcolor=blue,urlcolor=blue}

\DeclareGraphicsExtensions{.pdf, .png, .jpg}

\begin{document}%

\begin{abstract}
  A numerical study of the dynamic of the nonselfgravitating, unmagnetized, nonaxisymmetric,
  and rotating the torus around the non-rotating black hole is presented. We investigate
  the instability of the rotating torus subject to perturbations presented by increasing or
  decreasing the angular velocity of the stable torus. We have done, for the first time,
  an extensive analysis of the torus dynamic response  to the perturbation of the angular
  velocity of the stable torus. We show how the high, moderate, and low values of
  the perturbations affect the torus dynamic and help us to understand the properties of
  the instability and Quasi-Periodic Oscillation (QPO). Our numerical simulations indicate
  the  presence of Papaloizou-Pringle instability (PPI) with global $m=1$ mode
  and QPOs  for the moderate and lower values of the perturbations on the angular
  velocity of the stable torus. Furthermore, with the lower values of the perturbations,
  the torus can lead to a wiggling initially and then PPI is produced in it. Finally, the
  matter of the torus would be dissipated due to the presence of a strong torque.
\end{abstract}

% Keywords
%
\maketitle%
%

%%%%%%%%%%%%%%%%%%%%%%%%%%%%%%%%%%%%%%%%%%%%%%%%%%%%%%%%%%%%%%%%%%%%%%%
%%%%%%%%%%%%%%%%%%%%%%%%%%%%%%%%%%%%%%%%%%%%%%%%%%%%%%%%%%%%%%%%%%%%%%%
%%%%%%%%%%%%%%%%%%%%%%%%%%%%%%%%%%%%%%%%%%%%%%%%%%%%%%%%%%%%%%%%%%%%%%%
%%%%%%%%%%%%%%%%%%%%%%%%%%%%%%%%%%%%%%%%%%%%%%%%%%%%%%%%%%%%%%%%%%%%%%%
%%%%%%%%%%%%%%%%%%%%%%%%%%%%%%%%%%%%%%%%%%%%%%%%%%%%%%%%%%%%%%%%%%%%%%%
%%%%%%%%%%%%%%%%%%%%%%%%%%%%%%%%%%%%%%%%%%%%%%%%%%%%%%%%%%%%%%%%%%%%%%%

\section{Introduction}
\label{Introduction}

The accretion tori around the stellar and massive black holes are the one
of the interest to explain the observed high energies ($X$- and $\gamma$-rays)
in different astrophysical scenarios\citep{Lee1, Meszaros1} such as active
galactic nuclei which may form during
the collapse of super-massive stars \citep{Rees1, Shibata1}, coalescing of the black hole and
neutron star \citep{Rezzolla1, Kyutoku1}, etc.

The analytic and numerical investigations of the oscillatory modes of
accreated tori have been focused for a few decades. The oscillating relativistic tori
in a strong relativistic region were numerically developed by \cite{ZRF1, ZFRM} and
reference therein. The axisymmetric modes of high frequency quasi-periodic oscillations
(HF QPOs) were investigated around the non-rotating and rotating black holes using a fixed
spacetime matrix. In order to examine the disk oscillation under the
influence of the radial perturbation
and their nonlinear coupling with other modes,
\cite{Lee2} studied the numerical simulations of the relativistic torus around
the black hole implementing a pseudo potential. They found vertical epicyclic frequencies.
We have previously found \citep{OrhanQPO2} the numerical study of the dynamical
instability of the  torus on the equatorial plane due to
perturbation which was represented injecting gas from the outer boundary of the domain.
It was shown that the mass accretion rate in the perturbed
torus strongly depended on the cusp location and dynamical changes of the torus triggered
the PPI.
Recently, the evolution of the perturbed torus implemented by adding the subsonic velocity to radial,
vertical,  and diagonal velocities of the torus had been carried
out by \cite{Parthasarathy1}. It was seen
that the both $X$-ray and vertical epicyclic modes might be strongly excited in tori.

The long term oscillatory behavior of the black hole torus system is known as
runaway instability \citep{Abamowicz1} in case of axisymmetric perturbation
but  it is called PPI \citep{Papaloizou1}
if we apply a nonaxisymmetric perturbation on
the rotating torus around the black hole. It is seen in recent numerical simulations that
eventhough the runaway instability does not have a significant impact on dynamic of the
rotating torus \citep{Montero1},
the nonaxisymmetric perturbations on the rotating torus may have a critical impact on the
disk instability and in the variation of high energetic astrophysical observations \citep{Narayan1}.
Based on the above insight, the aim of this paper is to extend the understanding of the dynamical
instability of the relativistic torus in the case of the nonaxisymmetric perturbation
of the angular velocity of the stable torus carried out
by \cite{Orhan5}. All numerical
simulations had been done by perturbing the torus which had a constant
specific angular momentum rotating around the non-rotating black hole.

The PPI which created due to interactions of the propagation waves across the
corotating radius is mostly explained by nonaxisymmetric perturbation of the torus
around the black hole \citep{Papaloizou1}. The instability was occurring when the waves
inside the radius are transferring their energy to the waves outside of the corotation
radius \citep{Blaes1}. Besides, the emerging the hydrodynamical instabilities
were strongly depended
on the physical parameters of the system such as the torus size,
angular momentum of the black hole,
angular momentum of the torus, and the type of the instability \citep{DeVHaw, OrhanQPO2, Orhan5}.
The observational evidence of the PPI in Seyfert galaxy
$NGC$ $1068$ was found by \cite{Burillo1}.
The noncircular dynamics of the gas torus were seen. It enhanced
the amplitude of the instability and produced the lopsided morphology on the gas torus.
Hence, it was  believed that, the complex kinematic and  the lopsided morphology of the gas torus
could be signature of the PPI.

The power spectrums of the observational data taken by various ground and space
base detectors indicate the possibility of QPOs from the various astrophysical phenomena.
These QPOs are the result of the consistent mechanism which is varying significantly with time.
Studying the QPOs behavior of the tori around the black holes would help determine 
the physical properties of the black holes such as spin and mass \citep{SilWagOr}.
The frequencies numerically determined  from oscillating
tori around the black hole create certain
ratios which may be used to identify the properties of black holes. For example, the ratio
$3:2$ is suggested to be  a resonance between the fundamental frequency (orbital) and its overtones
in a strong gravitational region.

The paper is organized as  
follows: equations, models, and initial setups of the relativistic torus are
given in Section \ref{Formulation of Relativistic Hydrodynamics}.  
In Section \ref{Numerical Results}, the numerical results from our simulations
are given and discussed in detail. The perturbations which produce QPOs and
PPI are revealed.
The results found our numerical calculations are summarized in
Section \ref{Conclusion}.
The geometrized unit, $c = G =1$ and the space-time
signature $(-,+,+,+)$ are used throughout the paper.  

%%%%%%%%%%%%%%%%%%%%%%%%%%%%%%%%%%%%%%%%%%%%%%%%%%%%%%%%%%%%%%%%%%%%%%%
%%%%%%%%%%%%%%%%%%%%%%%%%%%%%%%%%%%%%%%%%%%%%%%%%%%%%%%%%%%%%%%%%%%%%%%
%%%%%%%%%%%%%%%%%%%%%%%%%%%%%%%%%%%%%%%%%%%%%%%%%%%%%%%%%%%%%%%%%%%%%%%
%%%%%%%%%%%%%%%%%%%%%%%%%%%%%%%%%%%%%%%%%%%%%%%%%%%%%%%%%%%%%%%%%%%%%%%

\section{Equations, Models, and Initial Setups}
\label{Formulation of Relativistic Hydrodynamics}
%%%%%%%%%%%%%%%%%%%%%%%%%%%%%%%%%%%%%%%%%%%%%%%%%%%%%%%%%%%%%%%%%%%%%%%

A numerical study of the nonselfgravitating, unmagnetized, nonaxisymmetric
and rotating  torus around the non-rotating black hole
is considered to model the instabilities and QPOs in case of the nonaxisymmetric
perturbation.  In order to understand the dynamics of the torus and instabilities created
due to the interaction of matter with black hole, we have numerically  solved the General
Relativistic Hydrodynamical (GRH) equations on equatorial plane  by using
the fixed Schwarzchild spacetime metric (see  \cite{Orhan, Orhan2} for details regarding
the conserved form of the hydrodynamical equations, all formulations and their
numerical solutions). The relation among the pressure, rest mass density, and internal energy
is defined using the perfect fluid equation of state $P = (\Gamma-1)\rho\epsilon$ with
$\Gamma = 4/3$. Initially a steady state accreated torus is produced using the appropriate values of
the highest density of the torus $\rho_c = 1.140$ x $10^{-4}$, the mass
ratio of the black hole-torus  $M_t/M_{BH}=0.1$, the polytropic constant $K=4.969$ x $10^{-2}$,
the constant specific angular momentum $\ell_{0}=3.80$, 
inner $r_{in}=4.57$ and outer, $r_{out}=15.889$, radii of  the torus, cusp location $r_{cusp}=4.57$,
and orbital period $t_{orb}=151.6$ at $r_c=8.35$ \citep{OrhanQPO2, OrhanQPO3, OrhanQPO4}.

The negligible values are used in the rest of the computational domain after setting up the initial
stable torus. They are: atmosphere density $\rho_{atm} = 10^{-8}\rho_c$, pressure $p_{atm} = 10^{-8}p_c$, 
radial velocity $v^{r}=0.0$, and angular velocity $v^{\phi}=0.0$ \citep{OrhanQPO2}.
The general form of the Schwarzchild metric  is
used to define the non-rotating  black hole 
at the center of computational domain. The inner and outer boundaries along the radial
direction are located at $r_{min}=2.8M$, inside the apparent horizon,
and $r_{max}=200M$, respectively. The angular direction goes
from $0$ and $2\pi$. The computation domain is defined on equatorial of the spherical coordinate
which covers the interval ($N_r$ $X$ $N_{\phi}$) $=$ ($3072$ $X$ $256$).
The outflow boundary condition is set up close to or far away from the black hole
to avoid the unwanted oscillations. The more details about initial setups as
well as boundaries  used in our numerical simulation can be found
in \cite{OrhanQPO2, OrhanQPO3, OrhanQPO4}.

The stable initial tori are perturbed by increasing or
decreasing  the angular velocity of the torus
everywhere to the computational domain in each
model for different cases,
seen in Table.\ref{table:Initial Models}. So we have a chance
to find out an instability and QPOs from
to the perturbed torus. The instabilities and QPOs might be
used to explain the observed $X-$rays from
different astrophysical phenomena.

\begin{table*}
%\scriptsize 
%\tiny
%\footnotesize 
\caption{Initial values of angular velocities used as a
  perturbation in each case.  From left to right in each case:
  First column represents the model names  and the second one is the
  angular ($v_n^{\phi}$)
velocities of the tori which are used 
as a perturbation in the first time step of the simulation.
The total times of the simulations vary from  models to models. 
 \label{table:Initial Models}}
\begin{center}
%\vspace*{-2ex}
  \begin{tabular}{cccccc}
    \hline  \hline
    \;\;\;\;\;\;\;\;\;\;\;\;\;\;\;\;\;\;\;\;\;Case I     &  & \;\;\;\;\;\;\;\;\;\;\;\;\;\;\;\;\;\;Case II  &  & \;\;\;\;\;\;\;\;\;\;\;\;Case III \\
    \hline
 Model & $v_n^{\phi}$ & Model & $v_n^{\phi}$ & Model & $v_n^{\phi}$ \\
\hline
$MH_1$  & $v^{\phi} + 0.2 v^{\phi}$  &   $MM_1$ &  $v^{\phi} + 0.1 v^{\phi}$  &  $ML_1$ &  $v^{\phi} + 0.05 v^{\phi}$  \\
$MH_2$  & $v^{\phi} + 0.3 v^{\phi}$  &   $MM_2$ &  $v^{\phi} + 0.15 v^{\phi}$ &  $ML_2$ &  $v^{\phi} + 0.01 v^{\phi}$  \\
$MH_3$  & $v^{\phi} - 0.3 v^{\phi}$  &   $MM_3$ &  $v^{\phi} + 0.18 v^{\phi}$ &  $ML_3$ &  $v^{\phi} + 0.001 v^{\phi}$  \\
        &                         &   $MM_4$  & $v^{\phi} - 0.1 v^{\phi}$  \\
\hline  \hline 
  \end{tabular}
\end{center}
%  \tablenotetext{}{}
%\vskip -0.8truecm
\end{table*}
%

%%%%%%%%%%%%%%%%%%%%%%%%%%%%%%%%%%%%%%%%%%%%%%%%%%%%%%%%%%%%%%%%%%%%%%%
%%%%%%%%%%%%%%%%%%%%%%%%%%%%%%%%%%%%%%%%%%%%%%%%%%%%%%%%%%%%%%%%%%%%%%%
%%%%%%%%%%%%%%%%%%%%%%%%%%%%%%%%%%%%%%%%%%%%%%%%%%%%%%%%%%%%%%%%%%%%%%%
%%%%%%%%%%%%%%%%%%%%%%%%%%%%%%%%%%%%%%%%%%%%%%%%%%%%%%%%%%%%%%%%%%%%%%%
\section{Numerical Results}
\label{Numerical Results}

%%%%%%%%%%%%%%%%%%%%%%%%%%%%%%%%%%%%%%%%%%%%%%%%%%%%%%%%%%%%%%%%%%%%%%%

The evolution of the perturbed torus shows a chaotic behavior.
In order to measure the instabilities produced during this chaotic behavior,
we measure the mass
accretion rate.  The mass accretion rate computed at the equatorial plane can be handled
by using the following expression

\begin{eqnarray}
\frac{dM}{dt} = \int_{0}^{2\pi}{\alpha \sqrt{\gamma} \rho u^rd\phi},
\label{MassAccRate}
\end{eqnarray}

\noindent
where $\rho$, $u^r$, $\alpha$, and $\gamma$ are the rest-mass density of the torus,
four-velocity along the radial coordinate, the lapse function, and
the determinant of the three-matrix, respectively.

The angular velocity perturbation can trigger the non-axisymmetric instability in the radial
direction on the torus and produce a restoring force. As a consequence,
the oscillation modes are produced
on the nonaxisymmetric disk around the black hole. To define this instability, we compute
the Fourier power of the density. The mode power allows us to compute the saturation point of
the instability and its growth rate. The detailed explanations of how to compute the power modes $m_1$,
$m_2$ etc. are given in \cite{OrhanQPO2, DeVHaw}. 

The simulations reported in this paper are performed for a long time at least $10$ to $25$
orbital periods to observe the instabilities and their persistent mechanism.
Otherwise, they run until the code
crushes. Thus we can measure the saturation, postsaturation, growth rates of the PPI, and
QPOs observed from the continues emission of the system.

%%%%%%%%%%%%%%%%%%%%%%%%%%%%%%%%%%%%%%%%%%%%%%%%%%%%%%%%%%%%%%%%%%%%%%%
\subsection{Perturbation of the Stable Torus: Case I}
\label{CaseI}

%%%%%%%%%%%%%%%%%%%%%%%%%%%%%%%%%%%%%%%%%%%%%%%%%%%%%%%%%%%%%%%%%%%%%%%%

\begin{figure}
\begin{center}

\vspace{0.5cm}

\includegraphics[width=8.5cm]{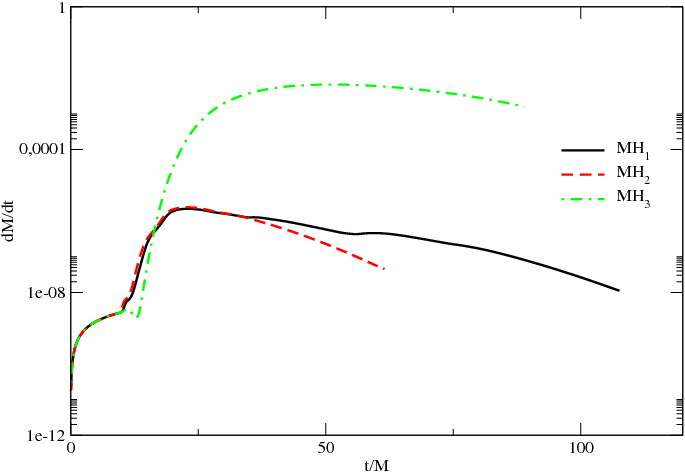}
\caption{The accretion rates computed from the perturbed torus around the
  non-rotating black hole plotted at the inner radius of the computational domain
  $r_{in}=2.8M$. The
  perturbation is represented by increasing or decreasing the angular velocity of the
stable torus, $20\%$ or more ( $Case \;I$).}
\label{CaseI_Mass_acc}
\end{center}
\end{figure}

\begin{figure}
\begin{center}

\vspace{0.5cm}

\includegraphics[width=8.5cm]{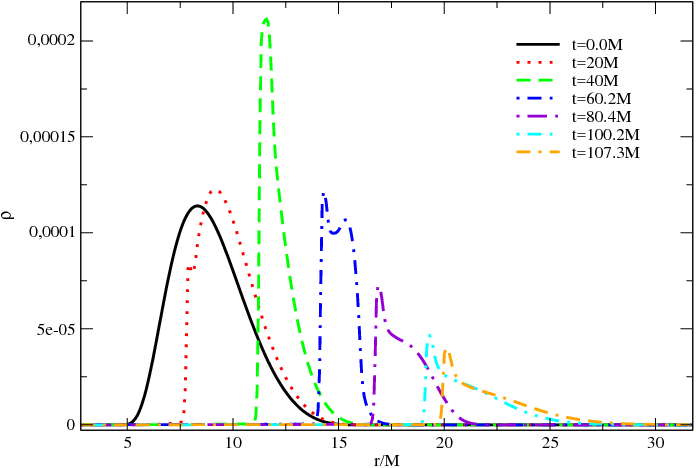}
\caption{Density of the perturbed torus at different snapshots. The perturbation
  is presented by increasing the angular velocity of the torus, $20\%$ (Model $MH_1$).}
\label{CaseI_density_diff_time}
\end{center}
\end{figure}

We have started by performing the nonaxisymmetric perturbation of the stable-rotating torus
around the non-rotating black hole using the high values of the angular velocity. Initially, we perturb
the torus by increasing or decreasing the angular velocity in an amount $20\%$ or more seen in
Table.\ref{table:Initial Models} and called $Case \;I$. Models $MH_1$ and $MH_2$ both represent the
perturbations in order to investigate the dynamical behavior of the torus by
increasing the angular velocity of the stable torus.
During the evolution, the angular momentum of the torus is transported inwards  through the unstable
point therefore less matter falls into the black hole as seen in Fig.\ref{CaseI_Mass_acc}.
It is also noted in Fig.\ref{CaseI_density_diff_time} that as a consequence of
increasing the angular velocity of the initial torus, the torus moves outward from the black hole and
tightens more  which causes an increase in the maximum density of
  initial stable torus. Later, the instability
is created and triggers the distribution of matter over the computational domain.
By contrast, Model $MH_3$ is constructed by decreasing the angular velocity of the torus
by amount $30\%$. As it is seen in Fig.\ref{CaseI_Mass_acc} that the angular momentum
is transported outward, more matter starts falling into the
black hole, and the torus is destroyed in less than a dynamical time step. The simulations in  $Case \;I$
do not produce any instability or show a prominent burst in a short time scale.
The rapid changes of the torus dynamic
in less than an orbital period produce an erratic behavior and it creates a strong shock. Therefore,
the code crushes.

%

%%%%%%%%%%%%%%%%%%%%%%%%%%%%%%%%%%%%%%%%%%%%%%%%%%%%%%%%%%%%%%%%%%%%%%%
%%%%%%%%%%%%%%%%%%%%%%%%%%%%%%%%%%%%%%%%%%%%%%%%%%%%%%%%%%%%%%%%%%%%%%%
\subsection{Perturbation of the Stable Torus: Case II}
\label{CaseII}

%%%%%%%%%%%%%%%%%%%%%%%%%%%%%%%%%%%%%%%%%%%%%%%%%%%%%%%%%%%%%%%%%%%%%%%%

\begin{figure}
\begin{center}
\vspace{0.5cm} 
\includegraphics[width=8.5cm]{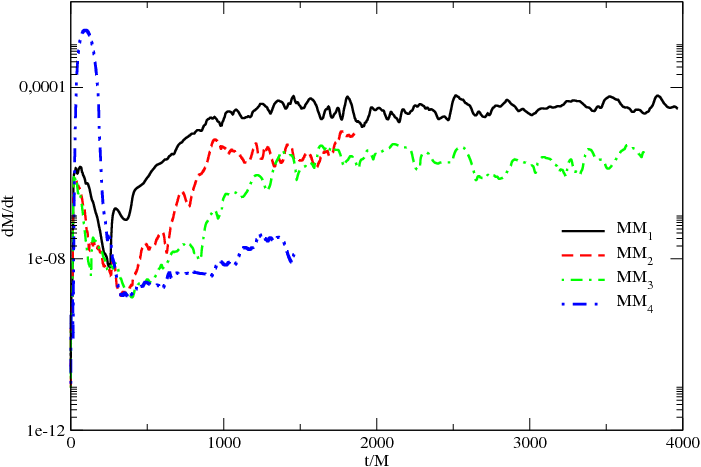}
\caption{The same as Fig.\ref{CaseI_Mass_acc} but it is for $Case \;II$.
  The torus is perturbed by increasing or decreasing the angular
  velocity of the stable torus with a moderate value,
  $20\% > v_n^{\phi} \geq 10\%$ (Case II).}
\label{CaseII_Mass_acc}
\end{center}
\end{figure}

\begin{figure}
\begin{center}
\vspace{0.5cm}
\includegraphics[width=8.5cm]{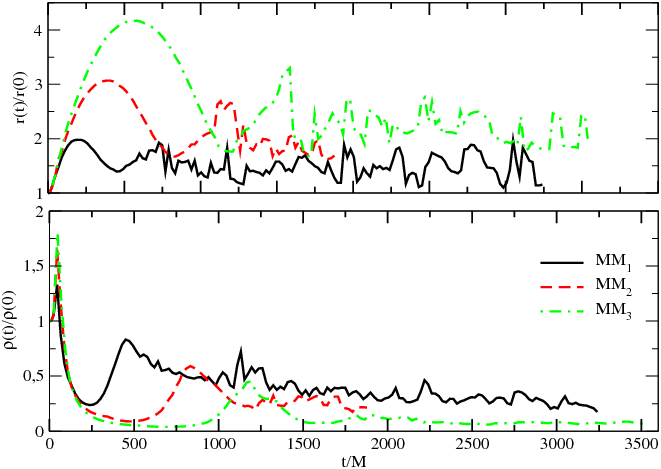}
\caption{The evolution of the torus for $Case \;II$. The
  radial location of the cusp (top panel)
  and the maximum density of the torus (bottom panel) are
  given. They are normalized  to their
  values at $t=0M$.}
\label{CaseII_Max_density}
\end{center}
\end{figure}

The oscillating the relativistic tori along the radial and angular directions  can
be handled by perturbing the torus angular velocity. In this section, we discuss the
numerical results in the presence of moderate amount of perturbation on angular
velocity of a rotating torus around a non-rotating black hole. Fig.\ref{CaseII_Mass_acc}
shows the evolution of the mass accretion rates computed at the inner boundary of the computational
domain $r=2.8M$ for  $Case \;II$ given in Table.\ref{table:Initial Models}. It is revealed in
this figure that
the instability grows initially due to a sudden change in angular momentum of the torus. It causes
the expansion  of the torus around the black hole and the cusp location moves outward. During this
expansion, $60M \leq t \leq 400M$, less amount of matter would be
accreated onto the black hole. Later, the torus starts falling toward to the black hole  and
the accreated mass increases exponentially. They reach their peak values  and then start to
oscillate approximately non-zero constant values for Models $MM_1$, $MM_2$, and $MM_3$ in $Case \;II$.
But the non-zero value could not be achieved by the initial perturbation given in Model $MM_4$.
These results imply that the moderate perturbation of the angular velocity,
$20\% > v_n^{\phi} \geq 10\%$, would trigger the instability with the non-zero constant values
of the physical parameters of the torus (i.e. the maximum density, the cusp location, etc.)
around the non-rotating black holes, seen in Fig.\ref{CaseII_Max_density}.   It is fair
the stressing in Fig.\ref{CaseII_Max_density} that  the maximum density and cusp locations for the Models
$MM_1$, $MM_2$, and $MM_3$ in $Case \;II$ survive with appreciable amplitudes after dynamics of the torus
reach to a quasi-steady state. So we expect to find
some PPI instabilities around the black holes.


\begin{figure}
\begin{center}
\vspace{0.5cm}  
\includegraphics[width=8.5cm]{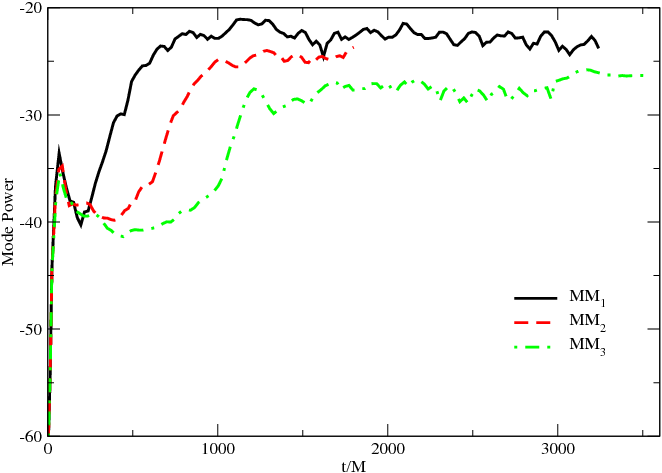}
\caption{The evolution of the $m=1$ mode amplitude for $Case \;II$.
  PPI is developed and it produces a quasi-periodic mode during
the evolution for all models, $MM_1$, $MM_2$ and $MM_3$.}
\label{CaseII_Mode_Power}
\end{center}
\end{figure}

Transporting the angular momentum of the torus outward can be activated by PPI
through the corotation point \citep{Narayan1}. It means that redistributing the angular
momentum of the torus
due to perturbation can trigger PPI. Fig.\ref{CaseII_Mode_Power} shows the mode power of the 
$m=1$ nonaxisymmetric structure  and saturation points for Models $MM_1$, $MM_2$ and $MM_3$.
We observe a saturation point in the early time of the simulation when $t \sim 70M$. It is created
due to the first kick on the torus. Later, the torus starts falling towards to the black hole and forms
a new cusp location and maximum density. The torus  becomes subject to PPI because the cusp is
located at a larger radius which can vary as seen in
Fig.\ref{CaseII_Max_density}. As shown in Fig.\ref{CaseII_Mode_Power},
after the saturation points are created around $t = 850M$,
$t = 1010M$, and $t = 1220M$ for Models $MM_1$, $MM_2$ and $MM_3$, respectively, the non-zero
values  of $m=1$  growth mode show a persistent structure during the evolution.
This mechanism is a clear
indication of PPI and leads to emission of $X-$ or $\gamma-$ rays in the observed black hole
torus systems, such as Seyfert galaxy  $NGC$ $1068$ \citep{Burillo1}.

\begin{figure*}
\begin{center}  
%\vspace{0.1cm}  
\includegraphics[width=6.0cm]{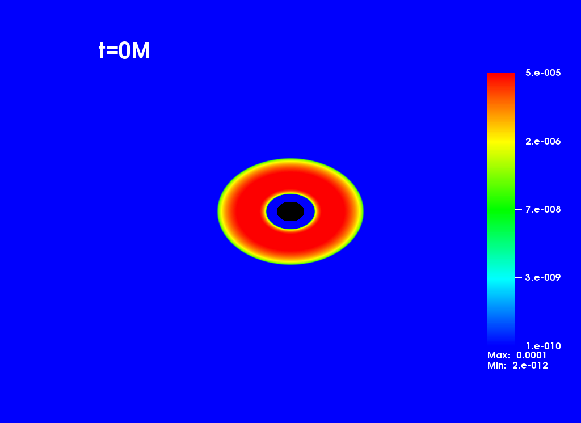}
\includegraphics[width=6.0cm]{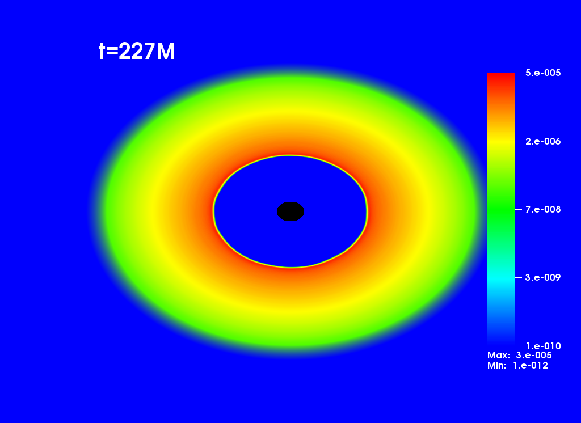}
\includegraphics[width=6.0cm]{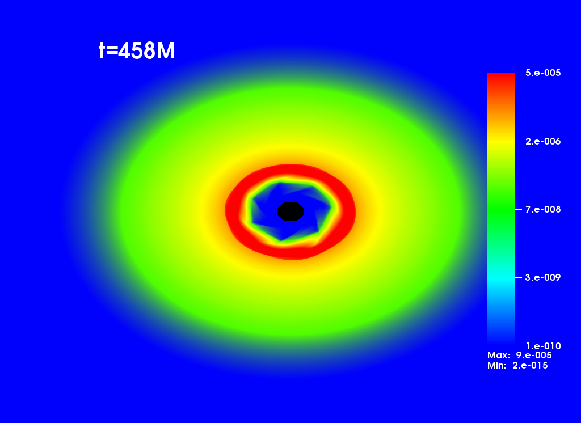}
\includegraphics[width=6.0cm]{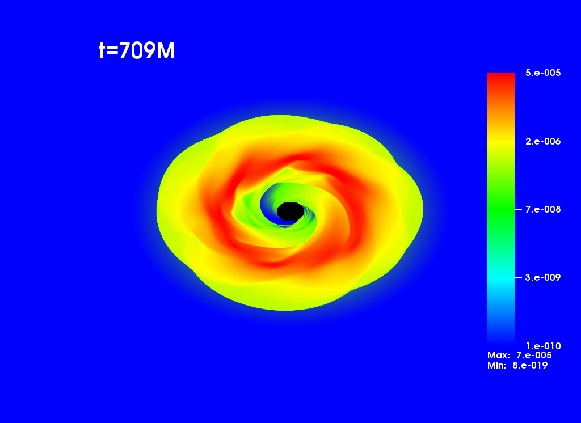}
\includegraphics[width=6.0cm]{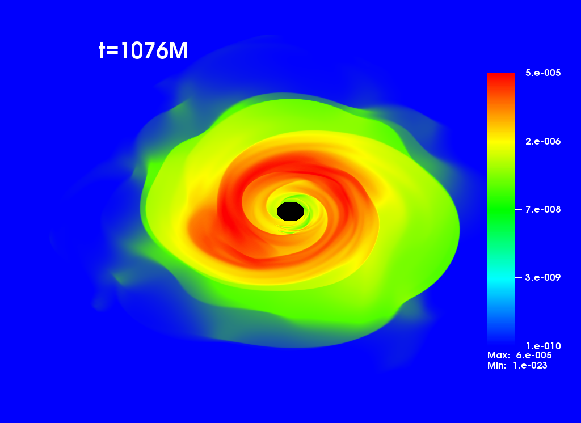}
\includegraphics[width=6.0cm]{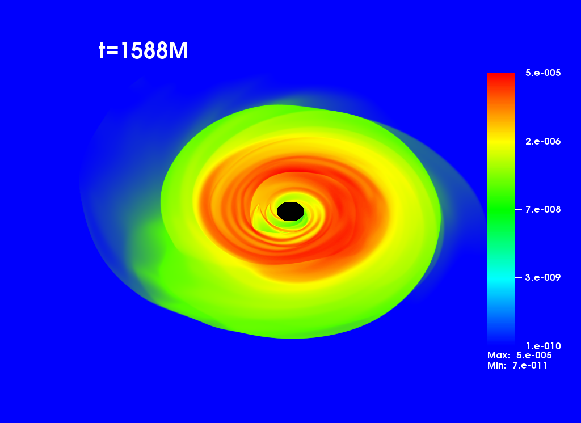}
\includegraphics[width=6.0cm]{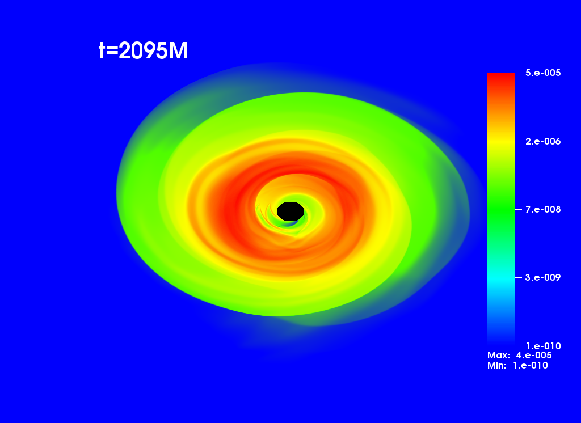}
\includegraphics[width=6.0cm]{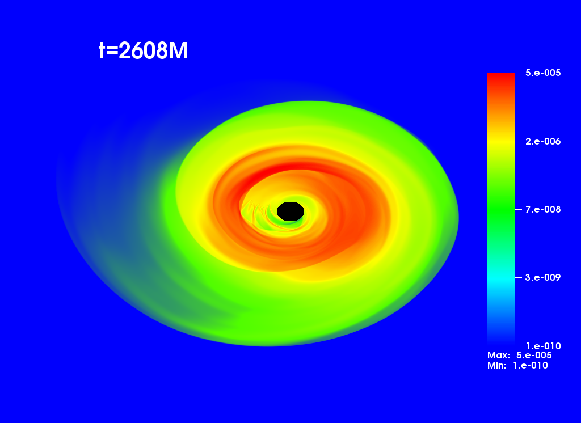}
\includegraphics[width=6.0cm]{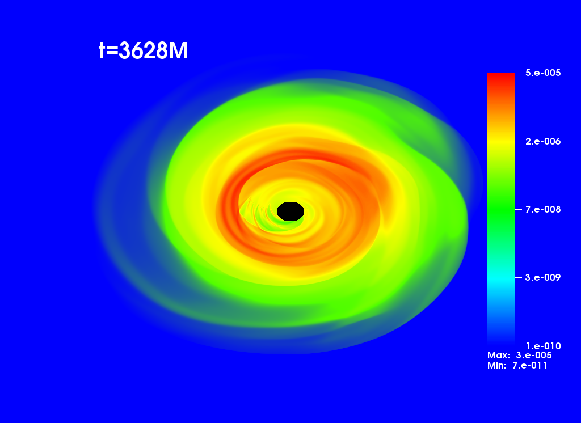}
\includegraphics[width=6.0cm]{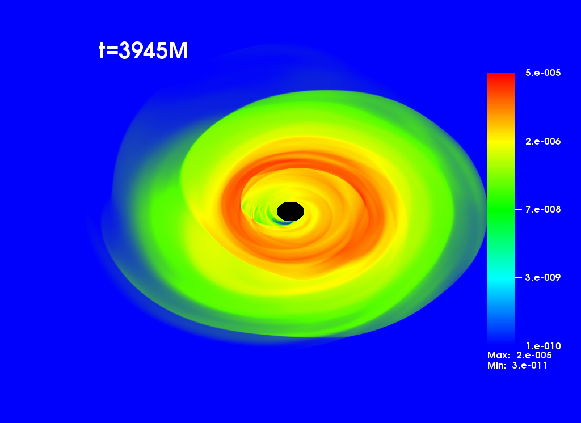}
\caption{The logarithmic change of the rest mass density at $10$
  different snapshots of the evolution for model $MM_1$ of $Case \;II$.
  The outer boundary of the domain seen in all panels is at
  $ [X_{min},Y_{min}] \rightarrow [X_{max},Y_{max}] =
  [-55M,-55M] \rightarrow [55M,55M]$  The spiral density
  waves  are formed after the quasi-rotation motion is
  appeared in the time evolution of the simulation.
So the growth of $m=1$ mode and the appearance of the PPI are noticeable.}
\label{CaseII_Different_snapshot}
\end{center}
\end{figure*}

Figure \ref{CaseII_Different_snapshot} shows ten different snapshots  of the logarithmic
rest-mass density of the torus  for Model  $MM_1$ of $Case \;II$.
The snapshot $t=0M$ shows the stable torus
dynamic and the others indicate the response of the torus dynamic to the perturbation.  It is seen in
Fig.  \ref{CaseII_Different_snapshot} that the cusp location of the torus moves away from the
black hole and the size of the torus  expands through the domain. Later, the torus
starts falling back toward to the black hole and forms a new cusp location which is almost $2$ times
bigger than the cusp location of the stable torus.  As noted from the snapshot $4$(four) at $t=709M$, the
matter falling back heats the gas and spiral shock wave is created. This spiral pattern creates a
persistent mechanics around the black hole, seen in the rests of the snapshots of
Fig.\ref{CaseII_Different_snapshot}. It is important to note that we can even see
the rotating spiral shock waves
around the black hole after $t=26$ orbital periods. The PPI and  $m=1$ mode structure survive for a long
time after the saturation of the PPI is reached.

\begin{figure}
\begin{center}
\vspace{0.5cm}  
\includegraphics[width=8.5cm]{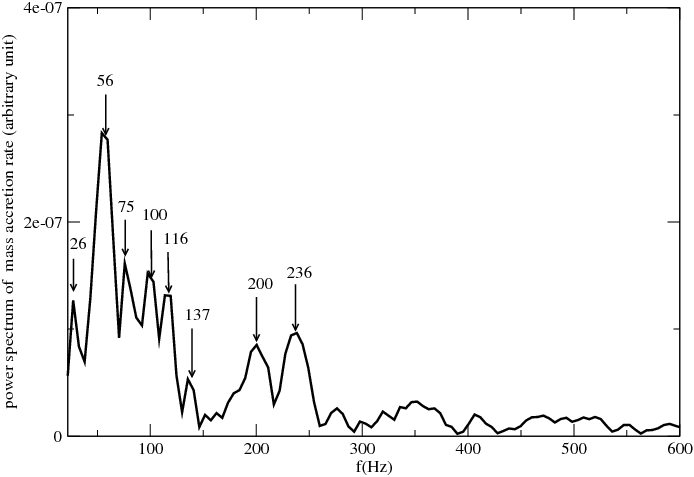}
\caption{The power spectrum of the mass accretion rate in
  arbitrary unit as a function of frequency for Model $MM_3$.
  The frequency axis is created assuming the black hole
  mass $M = 10M_{\odot}$. The fundamental frequency and overtones,
  which are results of the nonlinear
  couplings, are seen.}
\label{Case_II_QPOs}
\end{center}
\end{figure}

Formation of the QPOs around the black hole as a consequence of the torus-black hole interaction
due to any type of perturbation is a common phenomena \citep{OrhanQPO2,Orhan5} and these
QPOs can be used to define the physical properties of the black hole such as mass and
spin. In this paper, to reveal the QPO frequencies occurring in the torus around the black hole,
we compute the power spectrum of the mass accretion rate for  Model  $MM_3$, seen in
Fig.\ref{Case_II_QPOs}. It is indicated that the power law distribution of the oscillating torus
presents a fundamental frequency  $f=56 Hz$ and their overtones, $o_1=26 Hz$, $o_2=75 Hz$,
$o_3=100 Hz$, $o_4=116 Hz$, $o_5=200 Hz$, and  $o_2 = 236 Hz$. We see many harmonic
structures from the computed frequencies 1:2:3....

%

%%%%%%%%%%%%%%%%%%%%%%%%%%%%%%%%%%%%%%%%%%%%%%%%%%%%%%%%%%%%%%%%%%%%%%%
\subsection{Perturbation of the Stable Torus: Case III}
\label{CaseIII}

%%%%%%%%%%%%%%%%%%%%%%%%%%%%%%%%%%%%%%%%%%%%%%%%%%%%%%%%%%%%%%%%%%%%%%%%

\begin{figure}
\begin{center}
\vspace{0.5cm}  
\includegraphics[width=8.5cm]{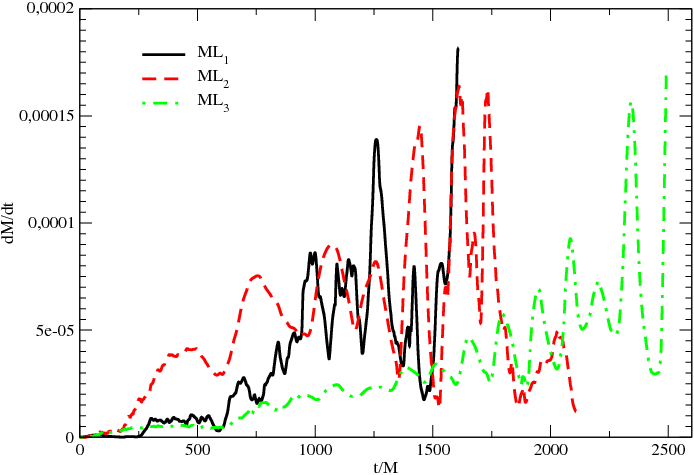}
\caption{The same as Fig.\ref{CaseI_Mass_acc} but it is for $Case \;III$. The torus is
  perturbed by increasing the angular velocity of the stable torus
with a small value, $5\% \geq v_n^{\phi} \geq 0.1\%$ ($Case \;III$).}
\label{CaseIII_Mass_acc}
\end{center}
\end{figure}

\begin{figure}
\begin{center}
\vspace{0.5cm}  
\includegraphics[width=8.5cm]{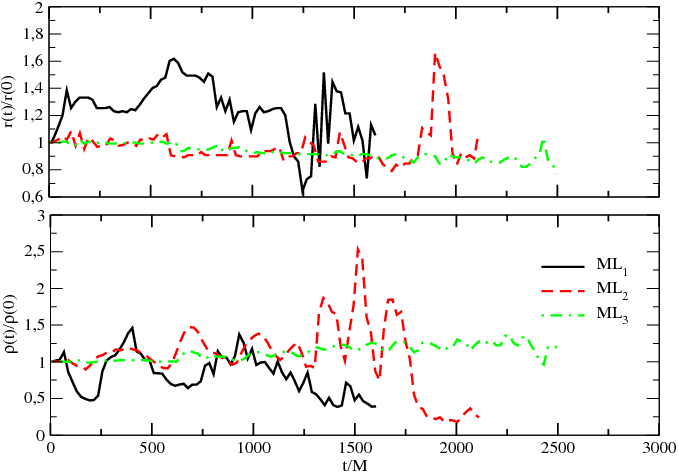}
\caption{The same as Fig.\ref{CaseII_Max_density} but for $Case \;III$.}
\label{CaseIII_Max_density}
\end{center}
\end{figure}

\begin{figure}
\begin{center}
\vspace{0.5cm}  
\includegraphics[width=8.5cm]{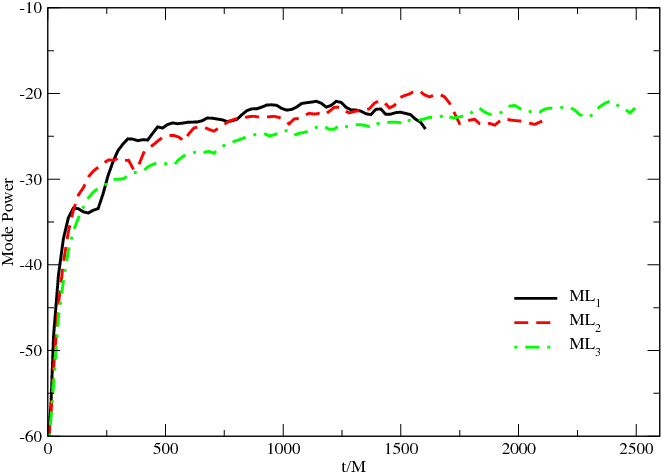}
\caption{The same as Fig.\ref{CaseII_Mode_Power} but it is
  for $Case \;III$. The low angular velocity
  perturbations create an instability in a way of different
  dynamical evolution. The produced
  instability grows very fastly after a few orbital periods
  and causes the code to crush.}
\label{CaseIII_Mode_Power}
\end{center}
\end{figure}

\begin{figure*}
\begin{center}  
\vspace{0.5cm}    
\includegraphics[width=6.0cm]{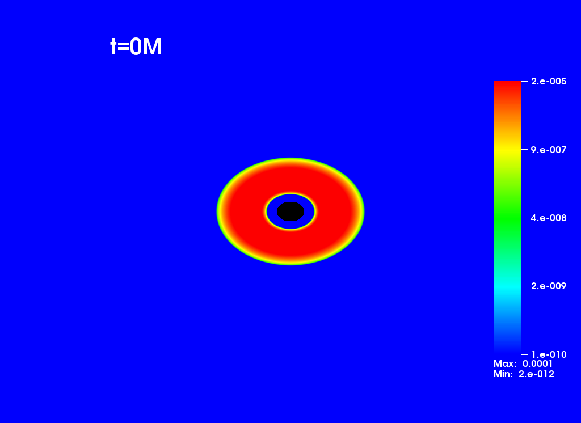}
\includegraphics[width=6.0cm]{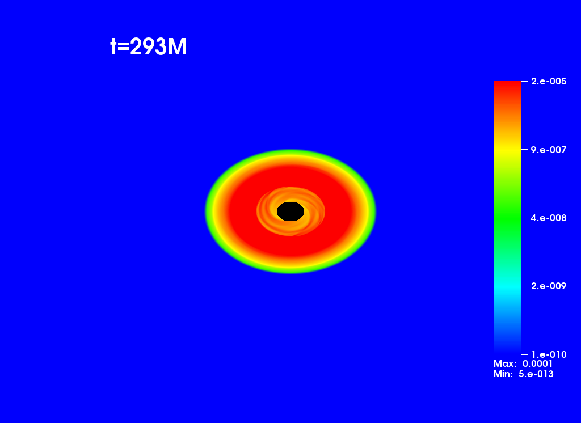}
\includegraphics[width=6.0cm]{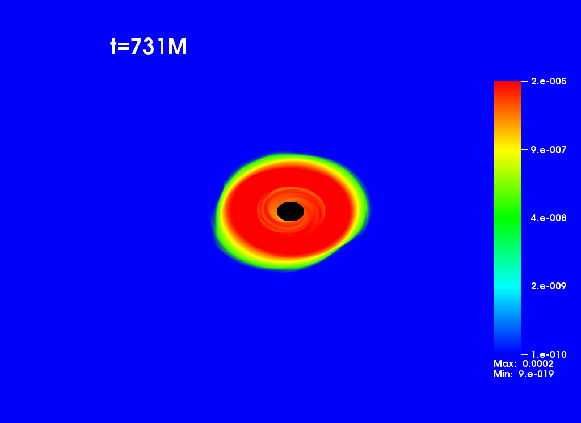}
\includegraphics[width=6.0cm]{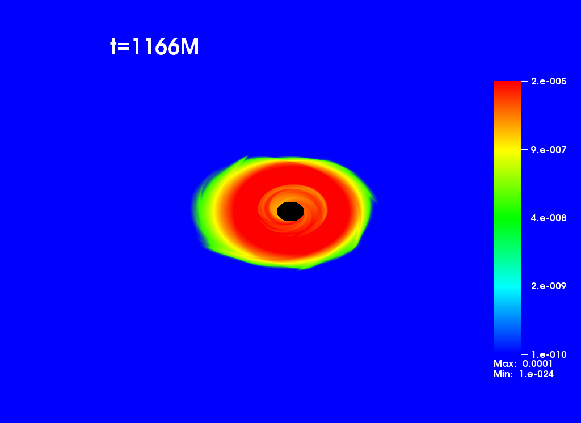}
\includegraphics[width=6.0cm]{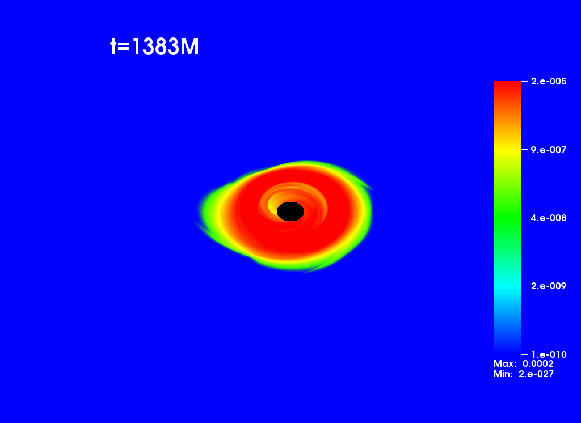}
\includegraphics[width=6.0cm]{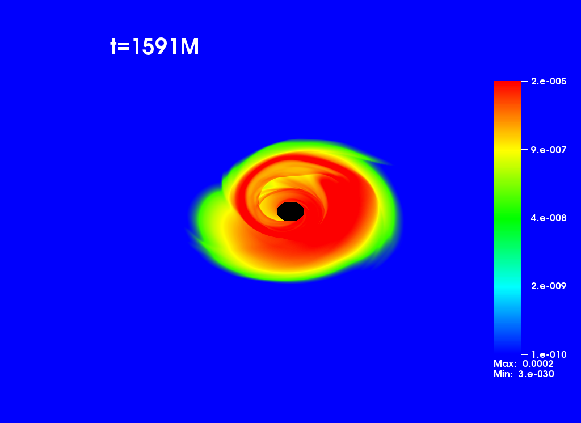}
\includegraphics[width=6.0cm]{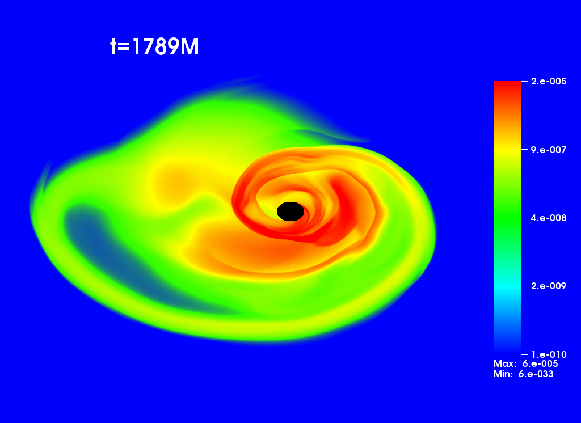}
\includegraphics[width=6.0cm]{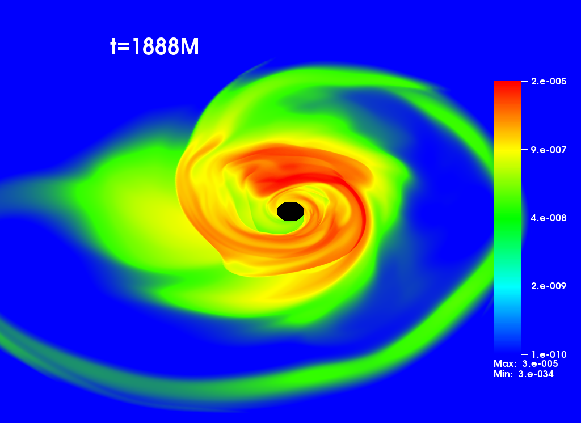}
\includegraphics[width=6.0cm]{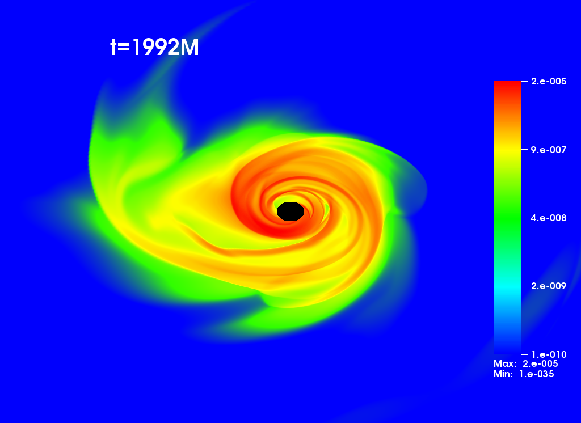}
\includegraphics[width=6.0cm]{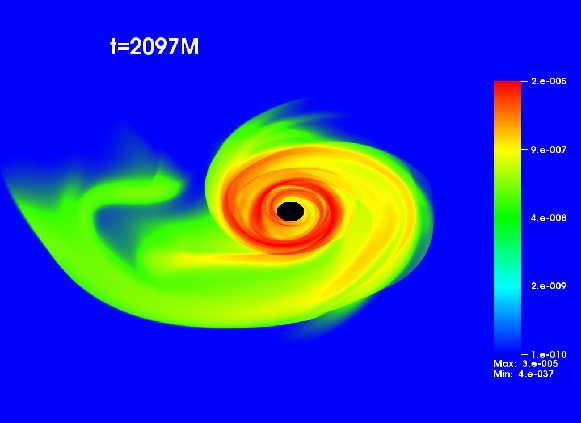}
\caption{The same as Fig.\ref{CaseIII_Different_snapshot} but for
  Model $ML_1$ in $Case \;III$.}
\label{CaseIII_Different_snapshot}
\end{center}
\end{figure*}

The results from representative models with the low values of angular velocity perturbation are shown
in Figs. \ref{CaseIII_Mass_acc} - \ref{CaseIII_Different_snapshot}. The mass accretion rates for all
models in $Case III$ represent the clear indication of PPI instability on the torus, seen in
Fig.\ref{CaseIII_Mass_acc}.  The small amount of perturbation on the angular velocity
of the torus triggers the instability but, as seen in
upper panel of Fig.\ref {CaseIII_Max_density} that, the amplitude of
the oscillation is too small initially due to pushing the torus cusp location either outward or inward
in a small amount.
The torus oscillates around a location of the cusp of the initial stable torus.
The torus  feels the initial kick even in the first time step and later it indicates some
wiggling in a quasi-periodic way (i.e. the torus starts to shake in the early time of the simulation).
Finally, the instability and the erratic behavior of the torus matter
are observed during the evolution, seen in Figs. \ref{CaseIII_Mass_acc}, \ref{CaseIII_Max_density},
and \ref{CaseIII_Different_snapshot}. These chaotic motion creates strong torques. During this
process, the spiral density waves, result of the growth of $m=1$, provide a channel with strong torque.
As seen in the lower panel of Fig.\ref{CaseIII_Max_density}, the matter falls into black hole  for
Models $ML_1$ and $ML_2$. On the other hand, it is noted in Fig.\ref{CaseIII_Max_density} that
Model $ML_3$ did not have enough time to create the
strong torque to destroy the disk dynamic and spiral structure. The recovery time of the instability
created on the torus is much larger in case of lower value  of the angular velocity
perturbation, shown in Fig.\ref{CaseIII_Mass_acc}.


The power mode analysis of the models given in $Case \;III$ shows that the growth of the PPI mode
appears after the saturation points are created. These points are varying for different models.
While the saturation points for Model $ML_1$ and $ML_2$ are created at $t=1228M$ and
$t=1560M$, respectively, seen in Fig.\ref{CaseIII_Mode_Power}, this point does not appear for
the Model $ML_3$ before the end of the simulation $t=2500M$ ($\sim$ orbital periods).
It is also clear in Fig.\ref{CaseIII_Mass_acc} that PPI growth coincides with a
significant oscillation in the mass accretion rate due to erratic behavior of the spiral shock
waves generated in the torus. In spite of very significant changes of the dynamic around the black hole,
PPI is not able to survive long enough  to create a persistent mechanism. Mode power starts to decrease
and it causes the code to crush due to strong shock generated inside the torus.

Fig.\ref{CaseIII_Different_snapshot} also shows the dynamical change  of the logarithmic rest-mass
density of the torus  at $t=0M$ (the stable torus)  and later times after the perturbation is applied,
for Model $ML_1$. The formation of the instability and inflowing spiral waves appear
in this figure. It is also seen how the spiral pattern spreads through the computation domain.
It  redistributes the disk material over the domain and is radially extended through the equatorial
plane.

%

%%%%%%%%%%%%%%%%%%%%%%%%%%%%%%%%%%%%%%%%%%%%%%%%%%%%%%%%%%%%%%%%%%%%%%%%
%%%%%%%%%%%%%%%%%%%%%%%%%%%%%%%%%%%%%%%%%%%%%%%%%%%%%%%%%%%%%%%%%%%%%%%%
%%%%%%%%%%%%%%%%%%%%%%%%%%%%%%%%%%%%%%%%%%%%%%%%%%%%%%%%%%%%%%%%%%%%%%%
%%%%%%%%%%%%%%%%%%%%%%%%%%%%%%%%%%%%%%%%%%%%%%%%%%%%%%%%%%%%%%%%%%%%%%%

\section{Conclusion}
\label{Conclusion}

%%%%%%%%%%%%%%%%%%%%%%%%%%%%%%%%%%%%%%%%%%%%%%%%%%%%%%%%%%%%%%%%%%%%%%%
%%%%%%%%%%%%%%%%%%%%%%%%%%%%%%%%%%%%%%%%%%%%%%%%%%%%%%%%%%%%%%%%%%%%%%%

We have numerically done an extensive analysis  of the oscillation properties of the
perturbed torus
in a strong general relativistic region  by applying the perturbation to the
angular velocity
of the initially stable torus. Any increase or decrease in the angular
velocity of the stable
torus triggers the stability properties due to the modified
centrifugal forces which
are dominant in the relativistic region close to the black hole.  

The nonaxisymmetric perturbation due to the sudden change of the angular
velocity of the stable torus in a high amount triggers the instability.
The torus starts to extend
outward if the the angular velocity increases. Otherwise,
driven instability can cause an increase in the mass flux exponentially and the
matter rotating around the black hole quickly falls into the black hole
and the torus would disappear less than in a dynamical timescale.
In the any case of the decreasing the initial angular velocity  of the stable
torus, the angular momentum
is transported outward, more matter starts falling into the
black hole, and the torus would be destroyed again  in less than an orbital period.

On the other hand, it is seen from our
numerical simulations that the increasing  in the angular velocity of the
stable torus in an amount $18\% \geq v_n^{\phi} \geq 10\%$
would lead to a PPI. The transporting of the angular momentum of the torus outward
can be activated by PPI through the corotation point. For the simulations in $Case \;II$,
which consist of the moderate values of the perturbation for the angular velocities,
the saturation points occur
at different times for various models and then the non-zero
values  of $m=1$  growth modes show a persistent structure during the evolution.
This mechanism is a clear indication of PPI and leads to an emission of $X-$ or
$\gamma-$ ray in the observed torus-black hole
system, such as Seyfert galaxy  $NGC$ $1068$ \citep{Burillo1}.
Our results also demonstrate that PPI produces a QPO in the torus-black hole
system. Exploring the QPOs allow us to estimate the black hole properties such as spin
and mass.

We have also revealed the presence of the PPI and QPOs in our simulations in case of
moderately low perturbation in the  angular velocity.
The torus  feels the initial kick even in the first time step and later it indicates some
wiggling in a quasi-periodic way.  Finally, the instability and the erratic behavior of the torus matter
are observed during the evolution.  These chaotic motions create a strong torque. During these
process, the spiral density waves result of the growth of $m=1$ provide a channel with a strong torque
which dissipates the matter of the torus. In the early times of the simulation the spiral wave has
a small amplitude that it does not indeed have a strong interaction with surrounding.
Once the amplitude of
the spiral wave becomes larger, the nonlinear behavior is seen around the black hole.
As a last, it is important to notice that the recovery time of the instability created
in the torus is much larger in case of lower
values  of the perturbations in the angular velocities.

\section*{Acknowledgments}
All simulations were performed using the Phoenix  High
Performance Computing facility at the American University of the Middle East
(AUM), Kuwait.\\

\end{document}